%% file: PIMRC2023.tex
\documentclass[conference]{IEEEtran}
\IEEEoverridecommandlockouts
\usepackage[LGR,T1]{fontenc}
\usepackage[latin9]{inputenc}
\usepackage{mathtools}
\usepackage{amsmath}
\usepackage{amssymb}
\usepackage{acronym}
\makeatletter
\usepackage{array}
\usepackage{units}
\usepackage{multirow}
\usepackage{xcolor}

\makeatother

\begin{document}
\title{A Carbon Tracking Model for Federated Learning: Impact of Quantization and Sparsification\\
\thanks{This paper is funded by the EU in the call HORIZON-HLTH-2022-STAYHLTH-01-two-stage under grant agreement No 101080564. The work is also supported by the Italian
National Recovery and Resilience Plan (NRRP) of NextGeneration EU,
partnership on "Telecommunications of the Future" (PE0000001 program
"RESTART").}}
\author{\IEEEauthorblockN{Luca Barbieri\textit{$^{1,2}$}, Stefano Savazzi\textit{$^{2}$}, 
Sanaz Kianoush\textit{$^{2}$},
 Monica Nicoli\textit{$^{1}$}, Luigi Serio\textit{$^{3}$}}
\IEEEauthorblockA{\textit{$^{1}$ Politecnico di Milano, Milan,
Italy}, \textit{$^{2}$ Consiglio Nazionale delle Ricerche, Milan, Italy}, \\
\textit{$^{3}$ Technology Department, CERN - 1211 Geneva 23 - Switzerland}
}
 }
\maketitle
\begin{abstract}
Federated Learning (FL) methods adopt efficient communication technologies to distribute machine learning tasks across edge devices, reducing the overhead in terms of data storage and computational complexity compared to centralized solutions.
Rather than moving large data volumes from producers (sensors, machines) to energy-hungry data centers, raising environmental concerns due to resource demands, FL provides an alternative solution to mitigate the energy demands of several learning tasks while enabling new Artificial Intelligence of Things (AIoT) applications. This paper proposes a framework for real-time monitoring of the energy and carbon footprint impacts of FL systems. The carbon tracking tool is evaluated for consensus (fully decentralized) and classical FL policies. For the first time, we present a quantitative evaluation of different computationally and communication efficient FL methods from the perspectives of energy consumption and carbon equivalent emissions, suggesting also general guidelines for energy-efficient design. Results indicate that consensus-driven FL implementations should be preferred for limiting carbon emissions when the energy efficiency of the communication is low (i.e., < 25 Kbit/Joule).  
Besides, quantization and sparsification operations are shown to strike a balance between learning performances and energy consumption, leading to sustainable FL designs. 
\end{abstract}

\begin{IEEEkeywords}
Federated Learning, Consensus, Energy Consumption, Green Machine Learning, Internet
of Things. 
\end{IEEEkeywords}

\IEEEpeerreviewmaketitle{}
\input{Sections/acronyms}

\input{Sections/Introduction}

\input{Sections/Section2}

\input{Sections/Section3}

\input{Sections/Section4}

\input{Sections/Conclusions}

\end{document}

%% file: Sections/acronyms.tex
\acrodef{WWAN}[WWAN]{Wide Wireless Area Network}

\acrodef{CNR}[CNR]{Consiglio Nazionale delle Ricerche}

\acrodef{ML}[ML]{Machine Learning}
\acrodef{AI}[AI]{Artificial Intelligence}
\acrodef{NN}[NN]{Neural Network}
\acrodef{CNN}[CNN]{Convolutional Neural Networks}

\acrodef{FL}[FL]{Federated Learning}
\acrodef{PS}[PS]{Parameter Server}
\acrodef{FedAvg}{FA][Federated Averaging}
\acrodef{CL}[CL]{Centralized Learning}
\acrodef{FD}[FD]{Fully-Decentralized}
\acrodef{CFA}[CFA]{Consensus-driven Federated Averaging}

\acrodef{IID}[IID]{Independent and Identical Distributed}
\acrodef{non-IID}[non-IID]{non Independent and Identical Distributed}

\acrodef{CE}[CE]{Cross Entropy}
\acrodef{Adam}[Adam]{Adaptive Moment Estimation}
\acrodef{SGD}[SGD]{Stochastic Gradient Descent}

\acrodef{MQTT}[MQTT]{Message Queuing Telemetry Transport}
\acrodef{IoT}[IoT]{Internet-of-Things}
\acrodef{QoS}[QoS]{Quality of Service}

\acrodef{RV}[RV]{Random Variable}
\acrodef{VMR}[VMR]{Variance to Mean Ratio}
\acrodef{D}[D]{Dispersion Index}

\acrodef{GPUs]}[GPUs]{Graphics Processing Units}

%% file: Sections/Introduction.tex
\section{Introduction}
Data centers are today a key component of many Artificial Intelligence of Things (AIoT) services, which rely on the network for data sharing and Artificial Intelligence (AI) for analytics.
They contribute 0.3 \% of the global equivalent
Green House Gas (GHG) emissions (and about $15\%$ of the emissions
of the entire Information and Communication Technology (ICT) ecosystem), which will further increase in the years
to come~\cite{kone1,tgcn}.
Federated Learning (FL)~\cite{drl} is emerging as a promising alternative to centralized AIoT, especially for training tasks where the data privacy needs to be protected (i.e., medical data): it distributes
the computing tasks across many edge devices possibly characterized by a more efficient use of the energy compared with data centers~\cite{first_look,tgcn}.
Combined with a judicious design of networking
and training stages, FL is expected to bring significant benefits
in terms of environmental impact, obviating in many cases the need
for a large centralized infrastructure for cooling or power delivery,
or reducing the emissions by migrating training tasks across different geographic locations according to time-dependent availability of sustainable energy. 

\begin{figure}
    \centering
    \includegraphics[width=\linewidth]{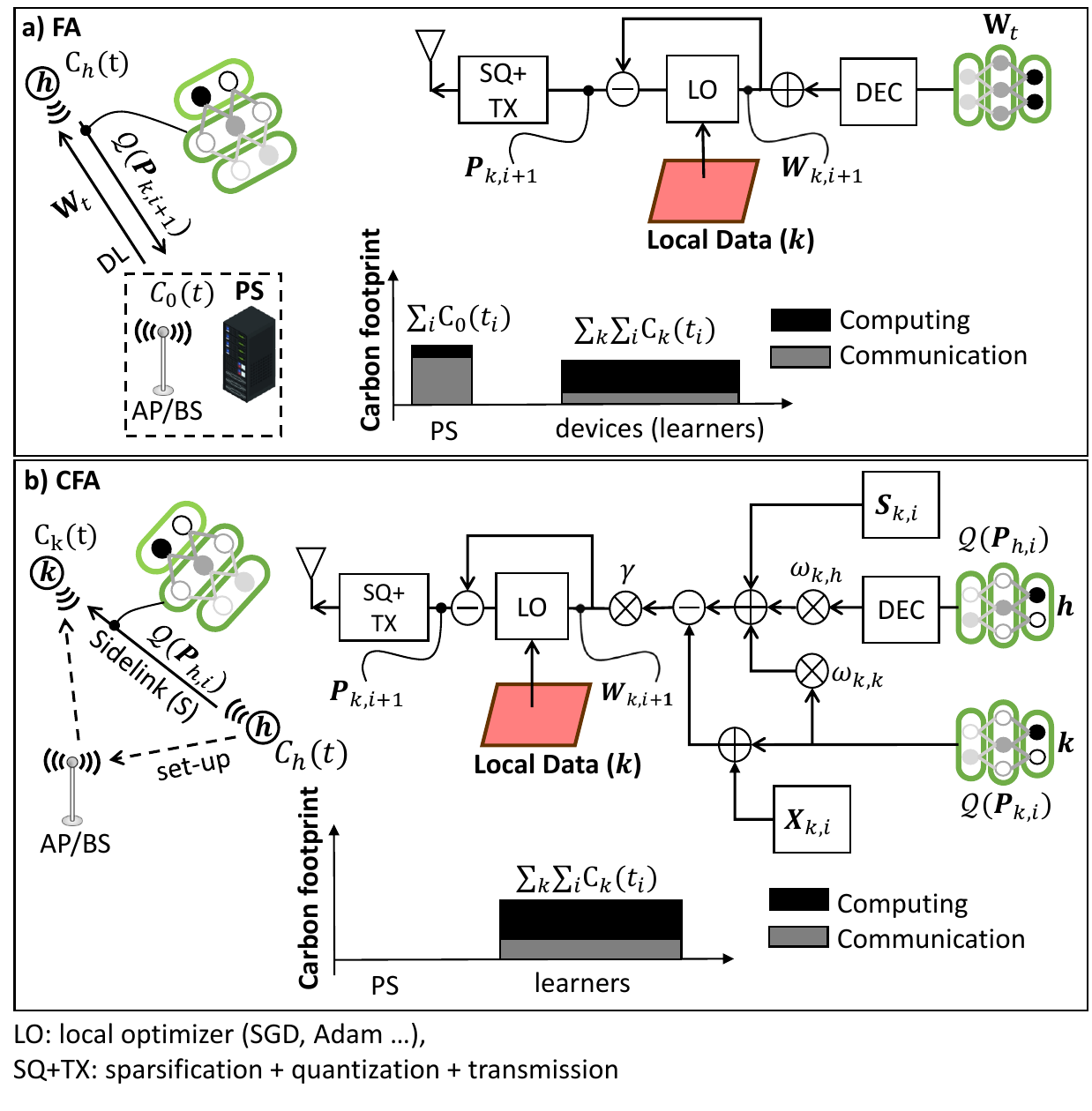}
    \vspace{-9mm}
    \caption{Federated Averaging (FA) relying on (a) Parameter Server (PS) for model aggregation and (b) Consensus process based on CHOCO-SGD tool. Sparsification and quantization operations and on-device carbon tracking.}
    \vspace{-4mm}
    \label{fig:cfa_vs_fa}
\end{figure}

FL architectures can be broadly divided into centralized and decentralized,
each involving different energy models, as shown in Fig.~\ref{fig:cfa_vs_fa}. For example, vanilla
Federated Averaging (FA)~\cite{drl} resorts to the classical server-client
architecture, where a Parameter Server (PS) coordinates the learning
process, collects the local models (rather than raw data as in classical
Machine Learning (ML)) and sends back the updated models (by averaging) to the devices. Both the PS and the federation of the devices contribute to the energy footprint.
On the other hand, in decentralized FL tools, such as Consensus-driven FA (CFA), the local model parameters
are shared and synchronized across multiple learners via
mesh, or Device-to-Device (D2D) networking, without relying on the
PS~\cite{commmag}. 

The problem of quantifying the energy and GHG
emissions, namely the carbon footprint, of FL has been recently tackled (see e.g.,~\cite{first_look,tgcn})
although the development of sustainable design and monitoring tools still remains
partially unexplored. Quantitative and qualitative evaluation of energy-efficient
FL strategies has been also carried out by comparing different implementations~\cite{comp,harvesting}. Quantization and sparsification
approaches~\cite{acsgd,qsgd} along with optimized strategies
for selecting suitable devices or informative ML model parameters~\cite{camad,layer_access} to be exchanged during the FL process
are to be considered as critical for reducing FL emissions.

\textbf{Contributions.} This paper develops a \emph{carbon tracking}
framework aiming to provide: 1) a systematic reporting of carbon and
energy footprints in FL processes over wireless networks; and 2) a toolkit to assess and compare different computationally
efficient methods for ML model parameter selection, compression and
quantization, adapted to both centralized (FA) and decentralized (CFA)
processes, and taking into account their carbon and energy footprints. 
Compared to the previous works~\cite{first_look,tgcn}, the developed tool quantifies explicitly the impact of model parameter sparsification and quantization, proposed in emerging FL designs \cite{acsgd,gossipc}, on the energy balance.
In particular, the proposed carbon tracking framework quantifies the carbon emissions on each FL round and accounts for the emissions
from energy grids, as well as the energy outputs from CPU/GPUs of
the individual devices. 
Results consider two increasingly complex image classification tasks, namely MNIST \cite{mnist} and CIFAR10 \cite{cifar10}, as well as ML model sizes to comprehensively evaluate the energy/carbon consumption of FA and CFA implementations. 
The goal is to study the optimal configuration of quantization and ML model sparsification to limit the FL process energy demands while maximizing learning performances. 
The analysis shows that decentralized FL (CFA) policies should be preferred when energy-inefficient communication protocols are adopted.
On the other hand, vanilla FA solutions may be selected when more efficient uplink/downlink communications are available.   
 Moreover, devices and PS carbon emissions need also to be well balanced to provide sustainable training processes. 

The remainder of this paper is organized as follows. 
Sec.~\ref{sec:Energy-footprint-modeling} introduces the carbon tracking framework, while Sec.~\ref{sec:compression_fl} presents the quantization and sparsification strategies adopted. 
Sec.~\ref{sec:Carbon-footprint-assessment} assesses the energy/carbon footprint of the proposed FL designs.
Finally, Sec.~\ref{sec:Conclusions-and-open} concludes the paper.

%% file: Sections/Section2.tex
\section{Carbon tracking framework \label{sec:Energy-footprint-modeling}}

Energy and carbon footprint models in FL (see~\cite{first_look,tgcn}
for a review) assume that the energy budget of the learning process
can be broken down into computing and communication costs. Notice
that both the PS (denoted with index $k=0$), when used, and the learners/devices ($k>0$)
contribute to the energy balance, and their footprints should be evaluated
separately. 
In what follows, we assume that a carbon
tracking update is issued on every new FL round $i$ that occurs at
(discrete) time $t_{i}\geq t_{0}$, with $t_{0}$ being the starting
time of the learning process. Let $\mathrm{C}_{k}(t_{i})$ refer to
the equivalent GHG emissions of device $k$ observed up to time $t_{i}$,
the proposed carbon tracking tool quantifies the GHG footprint through
an iterative approach 
\begin{equation}
\mathrm{C}_{k}(t_{i})=E_{k,i}\cdot I_{k,i}+\mathrm{C}_{k}(t_{i-1}),\label{eq:tracking_general-1}
\end{equation}
where $E_{k,i}$ is the energy cost to implement the new FL round
$i$, including local model optimization and communication, and $I_{k,i}=I_{k}(t_{i})$
is the Carbon Intensity (CI) of the electricity generation~\cite{LTE_power},
measured in Kg CO2-equivalent emissions per kWh (KgCO2-eq/kWh) at
time $t_{i}$. CI quantifies how much carbon emissions are produced
per kilowatt hour of locally generated electricity. The $I_{k,i}$
terms are typically monitored on an hourly basis~\cite{carbon_survey,tgcn}
as they depend on the specific regional energy grid where the device $k$ or PS is installed. Notice that grid differences can result
in large variations in eq. CO2 emissions~\cite{kone1}, as analyzed in Sec.~\ref{sec:Carbon-footprint-assessment}. 

In the following, we model the energy cost $E_{k,i}$ based on the
specific FL method employed. The communication costs are quantified
on average, in terms of the corresponding energy efficiencies ($\mathrm{EE}$),
standardized by the European Telecommunications Standards Institute (ETSI)~\cite{ee_etsi}. Efficiency terms in downlink (DL) $\mathrm{EE}_{\mathrm{D}}$,
uplink (UL) $\mathrm{EE}_{\mathrm{U}}$, or sidelink (S) $\mathrm{EE}_{\mathrm{S}}$
transmissions are measured here in bit/Joule {[}bit/J{]}~\cite{ee}.
The efficiencies also include the power dissipated in the RF front-end, baseband processing, and transceiver stages.

\subsection{Energy tracking for Parameter Server based FL (FA)}

In vanilla FL methods (FA), the PS collects
the local models produced by $K$ learners and produces a global model
of size $b_{\mathbf{W}}$ bits, which is fed back to active devices
on each round. Learners are powered on as they run the local optimizer
(LO) and decode the updated global model $\mathbf{W}_{i-1}$ obtained
from the PS at round $i-1$. During a FL round, the energy
cost $E_{k,i} = E_{k,i}^{(\mathrm{FA})}$ {[}Joule{]} of a device ($k>0$) can be written as 
\begin{equation}
E_{k,i}^{(\mathrm{FA})}=E_{k,i}^{(\mathrm{C})}+\frac{b_{\mathbf{W}}}{\mathrm{EE}_{\mathrm{D}}}+\frac{\mathbb{Q}_{k,i}[b_{\mathbf{W}}]}{\mathrm{EE}_{\mathrm{U}}}+E_{k,\mathbb{Q}}^{(\mathrm{C})}+E_{k,\mathrm{sleep}}^{(\mathrm{C})},\label{eq:fagen}
\end{equation}
namely, the superposition of the energy spent for receiving the global
model from the PS $\frac{b_{\mathbf{W}}}{\mathrm{EE}_{\mathrm{D}}}$,
the cost for the LO at round $i$ $E_{k,i}^{(\mathrm{C})}$ required for SGD computation, and the UL communication $\frac{\mathbb{Q}_{k}[b_{\mathbf{W}}]}{\mathrm{EE}_{\mathrm{U}}}$
of the selected local model parameters according to the model selection
and quantization policy $\mathbb{Q}_{k}[\cdot]$ (see Sec.~\ref{sec:compression_fl}). Implementation
of policy $\mathbb{Q}_{k}[\cdot]$ has a cost which is quantified here
as $E_{k,\mathbb{Q}}^{(\mathrm{C})}$. Finally, $E_{k,\mathrm{sleep}}^{(\mathrm{C})}$
is the energy cost in sleep mode, which is required by devices while
waiting for the global model to be produced by the PS. 

The PS ($k=0$)
energy per round is 
\begin{equation}
E_{0,i}^{(\mathrm{PS})}=K\cdot E_{0,\mathrm{global}}^{(\mathrm{C})}+\frac{b_{\mathbf{W}}}{\mathrm{EE}_{\mathrm{U}}}+\sum_{k=1}^{K}\frac{\mathbb{Q}_{k,i}[b_{\mathbf{W}}]}{\mathrm{EE}_{\mathrm{D}}}+E_{0,\mathrm{sleep}}^{(\mathrm{C})},
\end{equation}
which accounts for the cost for $K$ global model updates each with cost $E_{0,\mathrm{global}}^{(\mathrm{C})}$,
global model DL publication $ \frac{b_{\mathbf{W}}}{\mathrm{EE}_{\mathrm{U}}} $, and
collection of local model parameters $\frac{\mathbb{Q}_{k,i}[b_{\mathbf{W}}]}{\mathrm{EE}_{\mathrm{D}}}$
from the $K$ learners.
The total carbon emissions produced by FA can be then evaluated as 
\begin{equation}
    \mathrm{C}_{\text{tot}}^{\text{(FA)}} = \sum_{i} \sum_{k = 1}^{K} \mathrm{C}_k^{(\text{FA})}(t_{i}) + \sum_{i} \mathrm{C}^{(\text{PS})}_0(t_{i})
\end{equation}
where $\mathrm{C}^{(\text{FA})}_k(t_{i}) = E_{k,i}^{\text{(FA)}} \cdot I_{k,i} + \mathrm{C}_k^{(\text{FA})}(t_{i-1})$ is the carbon footprint of device $k$ at round $i$ while $\mathrm{C}_0^{(\text{PS})}(t_{i}) = E_{0,i}^{\text{(PS)}} \cdot I_{0,i} + \mathrm{C}_0^{(\text{PS})}(t_{i-1})$ is the carbon footprint of the PS at round $i$.

\subsection{Energy tracking for decentralized CFA}

Decentralized CFA techniques~\cite{drl,commmag} do not employ the
PS as the devices mutually exchange their local model parameters,
i.e., through publish-subscribe operations~\cite{access}. Each learner
is responsible for producing a global model representation which is
the result of a consensus over the received local model parameters
obtained from neighbor devices. CFA adopts a distributed
weighted averaging approach~\cite{commmag,drl} used to combine the
received neighbor models. 

As shown in Fig. \ref{fig:cfa_vs_fa}, the devices mutually exchange their local model parameters with an assigned number $N<K$ of neighbors. Let $\mathcal{N}_{k,i}$ be the set that contains the $N$ chosen neighbors of node $k$ at round $i$, the energy cost $E_{k,i}=E_{k,i}^{(\mathrm{CFA})}$ of an individual learner ($k>0$) can be written as 
\begin{equation}
\begin{aligned}E_{k,i}^{(\mathrm{CFA})}={} & E_{k,i}^{(\mathrm{C})}+N\cdot E_{k,\mathrm{global}}^{(\mathrm{C})}+\sum_{h\in\mathcal{N}_{k,i}}\frac{\mathbb{Q}_{h,i}[b_{\mathbf{W}}]}{\mathrm{EE}_{\mathrm{S}}}\\
{} & +\frac{\mathbb{Q}_{k,i}[b_{\mathbf{W}}]}{\mathrm{EE}_{\mathrm{S}}}+E_{k,\mathbb{Q}}^{(\mathrm{C})}+E_{k,\mathrm{sleep}}^{(\mathrm{C})},
\end{aligned}
\label{eq:cfa}
\end{equation}
where now each learner runs the LO (with cost $E_{k,i}^{(\mathrm{C})}$),
produces a global model representation via $N$ weighted averaging
steps $ E_{k,\mathrm{global}}^{(\mathrm{C})} $, distributes the local
(selected) parameters and obtains the neighbor ones using sidelink
communications. 

The total carbon consumption under CFA can be quantified as
\begin{equation}
\mathrm{C}_{\text{tot}}^{\text{(CFA)}} = \sum_{i}  \sum_{k = 1}^{K} \mathrm{C}^{(\text{CFA})}_k(t_{i}) , 
\end{equation}
with $C_k^{(\text{CFA})}(t_{i}) = E_{k,i}^{(\text{CFA})} \cdot I_{k,i} + C_k^{(\text{CFA})}(t_{i-1})$ being the carbon footprint of device $k$ at round $i$.

%% file: Sections/Section3.tex
\section{Quantization and parameter selection in FL}
\label{sec:compression_fl}

This section presents the compression strategies employed for evaluating
the impact of the communication on the energy/carbon footprint for
the proposed FL setups. To guarantee a fair comparison, we assume
that both centralized and decentralized FL tools rely on the same
compression operators. More specifically, we consider two widely-adopted
mechanisms, namely top-$t$ sparsification~\cite{topk} and probabilistic
quantization~\cite{qsgd}, which are applied independently by each
device member of the federation.

We consider a generic input vector $\mathbf{w},$ of $N_{P}$ elements $w_{n}$, $n=1,...,N_{P}$, that collects  the entries of the model parameters $\mathbf{W}_{k,i}$ or some surrogate quantities, such as the corresponding gradients or updates. The compression policy aims at obtaining
a lower bit representation of $\mathbf{w}$ by successively applying
sparsification and quantization as 
\begin{equation}
\bar{\mathbf{w}}=\mathcal{Q}(\mathbf{w})=f_{q}(f_{s}(\mathbf{w})),\label{eq:generic_quantization}
\end{equation}
where $\mathcal{Q}(.)$ denotes the overall compression function,
while $f_{s}(.)$ and $f_{q}(.)$ indicate the operators required
for top-$t$ sparsification and probabilistic quantization, respectively.
The sparsification operator $f_{s}(.)$ outputs a new representation
$\widetilde{\mathbf{w}}=f_{s}(\mathbf{w})$ that selects the $t$
largest absolute values of $\mathbf{w}$, and sets all other entries
to $0$. Quantization $f_{q}(.)$ encodes the sparsification output
$\widetilde{\mathbf{w}}$ via a randomized rounding operation. By
setting the output quantization bits to $N_{b}\leq N_{bc}$, with (typical)
$N_{bc}=32$ bits, the element $\bar{w}_{n}$ of $\bar{\mathbf{w}}$ is defined
as~\cite{qsgd} 
\begin{equation}
\bar{w}_{n}=\|\widetilde{\mathbf{w}}\|_{2}\cdot\mathrm{sign}(\widetilde{w}_{n})\cdot\xi_{n}(\widetilde{\mathbf{w}},2^{N_{b}}),
\end{equation}
where $\mathrm{sign}(.)$ denotes the sign operator, while $\xi_{n}(\widetilde{\mathbf{v}},2^{N_{b}})$ is defined as in~\cite{qsgd}.


Compression output $\bar{\mathbf{w}}=\mathcal{Q}(\mathbf{w})$ in
(\ref{eq:generic_quantization}) corresponds to the vector $\bar{\mathbf{w}}=[\bar{w}_{1} \cdots \bar{w}_{N_P}]^{\mathrm{T}}$, and
can be used as a lower bit representation of $\mathbf{w}$. The number
$\mathbb{Q}_{k,i}[b_{\mathbf{W}}]$ of bits that are sent by device
$k$ on round $i$ according to carbon tracking models (\ref{eq:fagen})-(\ref{eq:cfa}), can be quantified as 
\begin{equation}
\mathbb{Q}_{k,i}[b_{\mathbf{W}}]=\delta \cdot\dfrac{N_{b}}{N_{bc}}\cdot b_{\mathbf{W}}=t\cdot N_{b},\label{eq:bits_used}
\end{equation}
where $\delta = \tfrac{t}{N_{P}}$ represents the fraction of the model parameters
selected by top-$t$ sparsification function $f_{s}(.)$,
while $\tfrac{N_{b}}{N_{bc}}$ sets the quantization level according
to $f_{q}(.)$ for each parameter. 

In what follows, we review the specific operations required for implementing
the considered compression strategies for the FA (Sec.~\ref{subsec:compression_vanilla_FL})
and CFA (Sec.~\ref{subsec:compression_consensus_FL})
schemes.

\subsection{Parameter Server based FL (FA)}

\label{subsec:compression_vanilla_FL}

In FA, the devices send to the PS the model updates being more suited for compression~\cite{drl}. Given the model
parameters $\mathbf{W}_{k,i}$, the model updates are evaluated as 
\begin{equation}
\mathbf{P}_{k,i}=\mathbf{W}_{k,i+1/2}-\mathbf{W}_{k,i},\label{eq:model_update}
\end{equation}
where $\mathbf{W}_{k,i+1/2}$ denotes the parameters obtained after applying the LO for device $k$ at round $i$. Then, $\mathbf{P}_{k,i}$
is compressed as 
\begin{equation}
\bar{\mathbf{P}}_{k,i}=\mathcal{Q}(\mathbf{P}_{k,i}), \label{eq:compressed_model_update}
\end{equation}
with $\mathcal{Q}(.)$ as in \eqref{eq:generic_quantization} and
transmitted to the PS. The PS updates its
global model by aggregating the received contributions as
\begin{equation}
\mathbf{W}_{t+1}=\mathbf{W}_{t}+\sum_{k=1}^{N}\sigma_{k}\bar{\mathbf{P}}_{k,i},
\end{equation}
where $\sigma_{k}$ is a weighting factor chosen as in \cite{tgcn}. Note that the PS forwards back the updated global model uncompressed
as in Sec.~\ref{sec:Energy-footprint-modeling}. 

\subsection{CFA based on CHOCO-SGD framework}

\label{subsec:compression_consensus_FL}

The CFA process considered here relies on the CHOCO-SGD
algorithm proposed in~\cite{gossipc}. This scheme employs two additional
variables stored at each device for preserving the average of the model iterates across consecutive rounds and for controlling the noise introduced by the compression~\cite{gossipc}. In the same spirit as FA, each device $k$ compresses the model updates
in a manner similar to \eqref{eq:model_update}-\eqref{eq:compressed_model_update}
leading to 
\begin{equation}
\bar{\mathbf{P}}_{k,i}=\mathcal{Q}(\mathbf{P}_{k,i})=\mathcal{Q}(\mathbf{W}_{k,i+1/2}-\mathbf{X}_{k,i}),\label{eq:model_updates_consensus}
\end{equation}
where $\mathcal{Q}(.)$ is defined as in \eqref{eq:generic_quantization}
and $\mathbf{X}_{k,i}$ is a local variable, with $\mathbf{X}_{k,0} = \mathbf{0}$ for round $i = 0$. 
The compressed representation 
$\bar{\mathbf{P}}_{k,i}$
is then exchanged over the network.

Upon receiving the contributions from its neighbors, device $k$ updates $\mathbf{X}_{k,i}$ as
\begin{equation}
\mathbf{X}_{k,i+1}=\mathbf{X}_{k,i}+\bar{\mathbf{P}}_{k,i}
\end{equation}
and then uses the compressed models received from the neighboring devices to update an additional local variable $\mathbf{S}_{k,i}$, with $\mathbf{S}_{k,0} = \mathbf{0}$ at round $i = 0$, as
\begin{equation}
\mathbf{S}_{k,i+1}=\mathbf{S}_{k,i}+\sum_{j\in\mathcal{N}_{k,i}}\omega_{k,j}\bar{\mathbf{P}}_{j,i},
\end{equation}
where $\omega_{k,j}$ is the $(k,j)$-th entry of a symmetric doubly
stochastic matrix $\boldsymbol{\Omega}$.
Finally, each device $k$ updates its local model using the following update rule 
\begin{equation}
\mathbf{W}_{k,i+1}=\mathbf{W}_{k,i+1/2}+\gamma(\mathbf{S}_{k,i+1}-\mathbf{X}_{k,i+1}),
\end{equation}
where $0 < \gamma \leq 1$ is the consensus step-size.



%% file: Sections/Section4.tex
\section{Carbon footprint evaluation\label{sec:Carbon-footprint-assessment}}

\begin{table}[!tb]
	\renewcommand{\arraystretch}{1.3}
	\caption{Model and energy parameters for FA and CFA schemes}
	\label{tab:energy_params}
	\centering
	\begin{tabular}{c|c|c}
		 & \textbf{MNIST} & \textbf{CIFAR10} \\
		\hline 
		$N_P$ & 59500 & 28146954 \\
		$b_{\mathbf{W}}$ & 0.24 MB & 112.59 MB \\
		$\delta$ & 10\%, 50\%, 100\% & 10\%, 50\%, 100\% \\ 
		$N_b $ & 8, 16, 32 & 16, 24, 32 \\
		\hline 
		$E_{k,i}^{(C)}$  & 3.51 Joule & 5.53 KJoule \\
		$E_{k,\mathbb{Q}}^{(C)}$ & 0.04 - 0.14 Joule & 18.9 - 66.2 Joule\\
		$E_{k,sleep}^{(C)}$ & 0.12 Joule & 59.92 Joule \\
		$E_{k,global}^{(C)}$ & 0.06 Joule & 29.96 Joule\\
		$I_k$ & 0.449 KgC02-eq/kWh & 0.449 KgC02-eq/kWh \\
		\hline

		$E_{0,sleep}^{(C)}$ & 0.70 Joule & 1.10 KJoule\\
		$E_{0,global}^{(C)} $ & 0.24 Joule & 114.02 Joule\\ 
		$I_0$ & 0.449 KgC02-eq/kWh & 0.449 KgC02-eq/kWh \\

	\end{tabular}
 \vspace{-2mm}
\end{table}

This section discusses the main factors that are expected to steer
the choice between vanilla FA and CFA paradigms towards sustainable designs. The goal is to provide a first look into the impact of quantization and sparsification of model parameters on carbon emissions and learning accuracy, for varying communication energy efficiencies typically found in wireless communication systems. We consider two scenarios
with increasing model size and training dataset complexities. The
first case (Sec.~\ref{subsec:mnist}) focuses on the MNIST classification task~\cite{mnist}, while
the second (Sec.~\ref{subsec:cifar}) concentrates on the CIFAR10 learning problem~\cite{cifar10}. Energy
and carbon footprints are influenced by the PS and device hardware
configurations. Table \ref{tab:energy_params} summarizes the main parameters used by the carbon tracking tool under the two
scenarios. For the PS (FA), we used a CPU (Intel i7 8700K, $3.7$ GHz, $64$ GB, GPU not used). A realistic pool of $K=10$ resource-constrained FL learners is adopted, namely Jetson Nano boards based on a low-power CPU (ARM-Cortex-A57
and GPU 128-core Maxwell).
Their energy expenditure parameters when implementing the FA and CFA schemes reviewed in Sec. \ref{sec:compression_fl} are summarized in Table \ref{tab:energy_params}.
Each device is designed to track its carbon emissions independently as a function of the estimated CI $I_{k,i} = I_{k}$ which is assumed as constant for all FL rounds.  

In what follows, rather than choosing a specific communication or carbon emission setting, we consider a what-if analysis approach as proposed in~\cite{first_look, tgcn}. We thus quantify the achievable loss/accuracy of the proposed FL designs under the assumption of different DL/UL and SL efficiencies (setting $\mathrm{EE}_{\text{COM}} = \mathrm{EE}_{\text{D}} = \mathrm{EE}_{\text{U}} = \mathrm{EE}_{\text{S}}$) and CI $I_{k}$ (see https://app.electricitymaps.com/map for reference values).
To guarantee a fair comparison, we also assume that the PS collects the (compressed) model updates from all $K$ devices. Similarly, CFA uses (all) $N=K-1$ neighbors. 
Each device uses an SGD LO with learning rate $0.01$, momentum $0.9$, and batch size of $64$ examples chosen in accordance with the computational capabilities of the chosen learners.
For CFA, we set $\gamma = 0.01$ in all experiments.
Note that the actual emissions may be larger than the estimated ones depending on the specific devices and implementations. Therefore, in the following, we will highlight relative comparisons.

\subsection{Impact of sparsification, energy and carbon efficiencies}
\label{subsec:mnist}

\begin{figure}[!t]
    \centering
    \includegraphics[width=\linewidth]{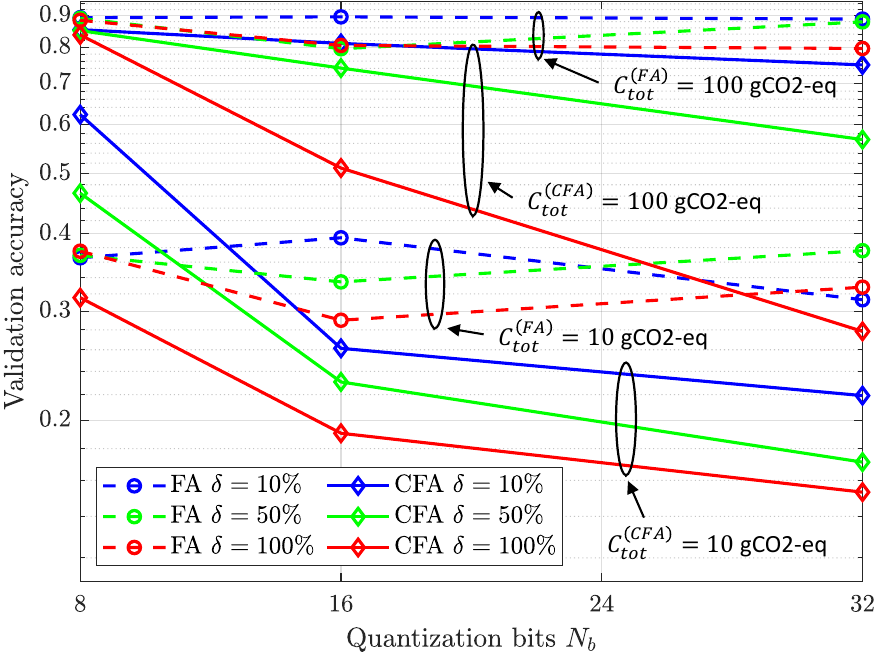}
    \vspace{-8mm}
    \caption{Analysis of the validation accuracy achieved by FA and CFA schemes under different levels of quantization and sparsification. Rings group together curves related to the same carbon footprint.}
    \label{fig:results_carbon_mnist}
    \vspace{-5mm}
\end{figure}

In the example, each device has access to $300$ observations randomly drawn from the MNIST database \cite{mnist}.
The devices employ a Lenet-5 model \cite{mnist}, with parameters in Table \ref{tab:energy_params}. 

Fig. \ref{fig:results_carbon_mnist} analyzes the quantization and sparsification impact on the learning accuracy by enforcing a max. total carbon emission (carbon budget) of $\mathrm{C}_{\text{tot}}$.
We consider two carbon emission targets, namely $\mathrm{C}_{\text{tot}} = \text{C}_{\text{tot}}^{(\text{FA})} = \text{C}_{\text{tot}}^{(\text{CFA})} = \{10,100\}$ gCO2-eq, modeling low to medium carbon consumptions. The quantization bits are in the range $N_b=8-32$ and the percentage of parameters shared $\delta=10\%-100\%$.
$\text{EE}_{\text{COM}} = 10$ Kbit/Joule, which roughly corresponds to an LTE design for macro-cell delivery \cite{ee, LTE_power}. The devices/PS are here located in the same region, i.e., Italy, with a corresponding CI $I_k = 0.449$ KgCO2-eq/kWh for $k = 0, \ldots, N$. 

Fig. \ref{fig:results_carbon_mnist} reports the validation accuracy obtained by the CFA and FA schemes under the two carbon budgets.
For stringent emission constraints, i.e., $C_{\text{tot}} = 10$ gCO2-eq, CFA provides better performances when $\delta = \{10\%, 50\%\}$ and $N_b = 8$ bits, while for $C_{\text{tot}} = 100$ gCO2-eq FA achieves higher accuracy regardless of the number of parameters shared and the quantization bits employed.  
ML model compression is more critical in CFA rather than in FA since it reduces the sidelink channel use. FA is not that much affected by the specific choice of the compression parameters as long as the carbon budget is high enough, i.e., $C_{\text{tot}} = 100$ gCO2-eq. On the other hand, CFA schemes must generally employ a more aggressive compression, i.e., $\delta = 10\%$ and $N_b = 8$ bits, to guarantee a reasonable target accuracy.
For FA, a good choice for the compression parameters is $\delta = 10\%$ and $N_b = 16$ bits as achieving the highest accuracy under all carbon budgets.

\begin{figure}[!t]
    \centering
    \includegraphics[width=\linewidth]{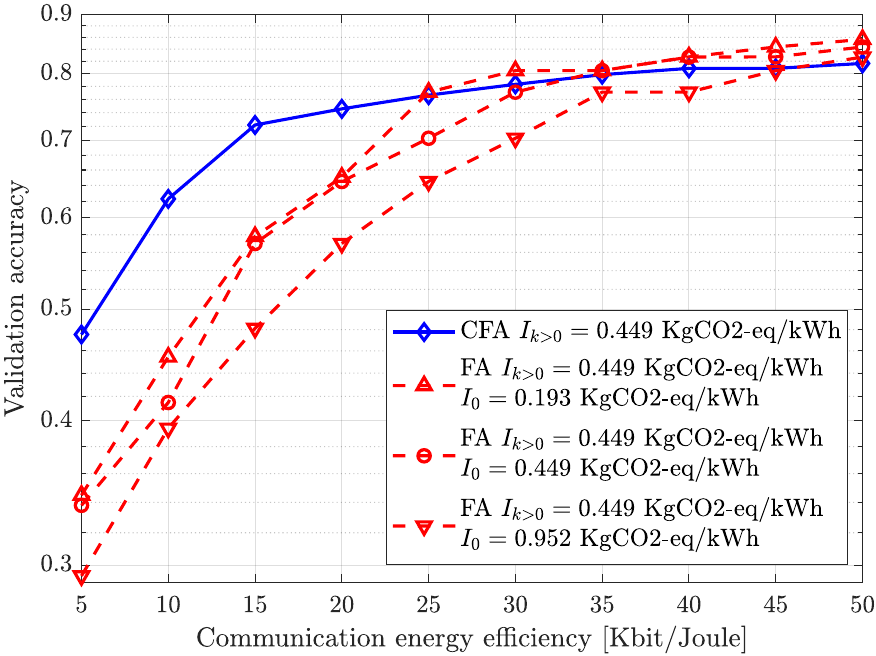}
    \vspace{-7mm}
    \caption{Analysis of the validation accuracy under carbon constraints with different communication (energy) efficiencies as well as carbon intensities.}
    \label{fig:results_comm_efficiency}
    \vspace{-4mm}
\end{figure}

Fig. \ref{fig:results_comm_efficiency} analyzes the impact of the communication energy efficiency as well as the CI on the learning accuracy under the same carbon budget $\mathrm{C}_{\text{tot}} = 10$ gCO2-eq. The energy efficiency $\text{EE}_{\text{COM}}$ varies from $5$ Kbit/Joule up to $50$ Kbit/Joule, with the last case corresponding to 5G micro-cell delivery and WiFi.
The results consider $\delta = 10\%$ and $N_b = 16$ bits for FA, while $\delta = 10\%$ and $N_b = 8$ bits for CFA following the previous analysis. 
In the same figure, we also study how the PS location and its CI $I_{0}$ affect the carbon consumption. 
Three CI values are considered, namely $I_0 = \{0.193, 0.449, 0.952\}$ KgCO2-eq/kWh, modeling different energy grid efficiencies. These values correspond to the geographical regions of Finland, Italy, and Poland, respectively. 
Comparing the results, CFA is shown to outperform FA for all CI terms, provided that the energy efficiency is below $25$ Kbit/Joule. 
On the other hand, when the communication is more efficient, centralized FA should be preferred. 
High carbon intensities, i.e., $I_0 = 0.952$ KgCO2-eq/kWh, make FA tools more susceptible to accuracy losses due to the high energy costs.   
Interestingly, even when the PS is located in the same region of the devices, CFA strategies are to be preferred as more accurate compared to PS-based solutions.    

\begin{table*}[!tb]
	\renewcommand{\arraystretch}{1.3}
 \vspace{-2mm}
	\caption{CIFAR10 analysis considering different carbon consumption targets, CI values and communication energy efficiencies. }
	\label{tab:cifar_results}
	\centering
	\begin{tabular}{c|c|c|c|c|c|c|c|c|c}
		\multirow{3}{*}{\textbf{Scenario}} & \multirow{2}{*}{$\delta$} & \multirow{2}{*}{$N_b$} & \multirow{2}{*}{$\text{EE}_{\text{COM}} $} & \multirow{2}{*}{$I_{0}$} & \multirow{2}{*}{$I_{k > 0}$} & \multicolumn{2}{c|}{Validation accuracy} & \multicolumn{2}{c}{Carbon}\\
		
		& & & & & & FA & CFA & $\text{C}_{\text{tot}}^{\text{(FA)}}$ & $\text{C}_{\text{tot}}^{\text{(CFA)}}$  \\
		
		 & [\%] & [bits] & [Kbit/Joule] & 
		 [KgCO2-eq/kWh] & [KgCO2-eq/kWh] & [\%] & [\%] & [KgCO2-eq] & [KgCO2-eq] \\
		\hline 
        
        \multirow{3}{*}{1} & 10 & 16 & 10 - 100 & 0.449 & 0.449 & \textcolor{green}{45.8} - \textcolor{green}{74.8} & 54.8 - 64.9 & 1.5 & 1.5 \\
        & 50 & 24 & 10 - 100 & 0.449 & 0.449 & 26.1 - 74.5 & \textcolor{green}{59.8} - \textcolor{green}{69.8} & 1.5 & 1.5 \\
        & 100 & 32 & 10 - 100 & 0.449 & 0.449 & \textcolor{red}{17.4} - \textcolor{red}{73.1} & \textcolor{red}{52.4} - \textcolor{red}{66.3} & 1.5 & 1.5 \\
        \hline  
        
        \multirow{3}{*}{2} & 10 & 16 & 10 - 100 & 0.449 & 0.449 & 70 & 70 & \textcolor{green}{3.6} - \textcolor{green}{0.6} & 15.6 - 4.2 \\
         & 50 & 24 & 10 - 100 & 0.449 & 0.449 & 70 & 70 & 5.1 -	0.7 & \textcolor{green}{14.8} - \textcolor{green}{1.8} \\
         & 100 & 32 & 10 - 100 & 0.449 & 0.449 & 70 & 70 & \textcolor{red}{8.6} - \textcolor{red}{1.1} & \textcolor{red}{27.3} - \textcolor{red}{2.9} \\
        \hline 
                      		
		\multirow{2}{*}{1} & 10 & 16 & 10 & 0.449 & 0.193 - 0.952 & \textcolor{green}{60.8} - 45.8 & 54.8 - 54.8  & 1.5 & 1.5 \\
		& 50 & 24 & 10 & 0.449 & 0.193 - 0.952 & \textcolor{red}{51.4} - \textcolor{red}{26.1} & 59.8 - \textcolor{green}{59.8} & 1.5 & 1.5 \\
		\hline 
		
		\multirow{2}{*}{2}  & 10 & 16 & 10 & 0.449 & 0.193 - 0.952 & 70 & 70 & \textcolor{green}{3.3} - \textcolor{green}{4.1} & \textcolor{red}{15.6} - \textcolor{red}{15.6} \\
		& 50 & 24 & 10 & 0.449 & 0.193 - 0.952 & 70 &  70 & 4.3 - 6.8 & 14.8 - 14.8 \\
		\hline 
		
	\end{tabular}
 \vspace{-2mm}
\end{table*}

\subsection{CIFAR10 analysis: impact of large model size}
\label{subsec:cifar}

To finalize the analysis, we here consider a more challenging image recognition task based on the CIFAR10 dataset and a much larger ML model. The goal is to evaluate the energy/carbon footprint of FA and CFA in a complex and potentially energy-hungry ML setup.
Each learner is assigned 500 examples for each one of the 10 classes and employs a VGG11 architecture \cite{vgg} with parameters defined in Table \ref{tab:energy_params}.

Table \ref{tab:cifar_results} summarizes the results of all the tests. In line with the previous analysis, we consider two main scenarios: 1) the learning process is constrained to a maximum carbon emission of $\text{C}_{\text{tot}}^{(\text{FA})} = \text{C}_{\text{tot}}^{(\text{CFA})} = 1.5$ KgCO2-eq and 2) the same process must achieve a $70$\% target accuracy with no constraints on carbon emissions. 
In particular, the first three rows refer to scenario 1 and show the validation accuracy obtained by FA and CFA tools for three quantization and sparsification cases, namely $(\delta, N_b) = (10\%, 16)$, $(\delta, N_b) = (50\%, 24)$ and $(\delta, N_b) = (100\%, 32)$, modeling moderate to no compression. We consider two communication efficiencies  $\text{EE}_{\text{COM}} = 10 $ Kbit/Joule and $\text{EE}_{\text{COM}} = 100 $ Kbit/Joule. 
The following three rows consider scenario 2 with same values for compression and energy efficiencies, and show now the carbon emissions required to reach $70$\% accuracy.
Comparing the results, in accordance with the analysis in Fig. 3, CFA provides a higher accuracy w.r.t. to FA schemes when the communication is inefficient, i.e.,  $\text{EE}_{\text{COM}} = 10 $ Kbit/Joule. FA is more effective when the learning process is implemented over more communication-efficient networks. 
Focusing now on the carbon emissions needed to reach a $70\%$ target accuracy, CFA requires a much larger footprint (i.e., $3 \times$) compared to FA for all cases considered.
Nevertheless, optimizing the compression parameters is beneficial for reducing carbon consumption with respect to uncompressed communications: for FA selecting $\delta = 10\%$ and $N_b = 16 $ bits allows to reduce the carbon consumption by roughly $50\%$ while for CFA $\delta = 50\%$ and $N_b = 24$ bits leads to a $42\%$ carbon reduction.

The last four rows of Table \ref{tab:cifar_results} analyze the impact of the regional CI on the performances and energy consumption with same configurations of the PS as described in the previous section. 
The results confirm the findings of Fig. \ref{fig:results_comm_efficiency}: FA tools are advantageous compared to CFA policies when the PS is located in a region with CI comparable to or better than the one of the devices.
Indeed, optimizing the PS location allows to further reduce the carbon consumption by $10\%$.


%% file: Sections/Conclusions.tex
\section{Conclusions\label{sec:Conclusions-and-open}}

This paper proposed a carbon tracking framework for monitoring the energy/carbon footprints of FL policies.
The proposed framework enables to assess the impact of ML model quantization and sparsification on the carbon/energy consumption for both centralized (FA) and consensus-driven (CFA) FL implementations. 
The developed tool keeps track of the energy/carbon demands in an iterative manner and accounts for computing/communication energy consumption as well as emissions arising from energy grids, allowing to identify the optimal operating conditions of FL processes and their energy expenditure. 
Experimental results are based on two increasingly complex image classification tasks and ML model sizes that serve as reference examples to study the optimal configuration of the compression parameters. 

The analysis shows that CFA tools are more suited under energy-demanding communication protocols (i.e., when the communication energy efficiency is below $25$ Kbit/Joule), while FA policies should be preferred under higher energy efficiency regimes. 
The optimization of quantization and sparsification operations in FA and CFA tools allows to further reduce the carbon footprint on average by roughly 50\% and 42\%, respectively, with respect to sending the ML model uncompressed. 
Besides, FA strategies have been shown to provide significant energy savings (up to 10\%) when the PS is located in a region with a carbon efficiency (intensity) that is at least $2$ times larger than that experienced by the devices. 

Further research activities may target the integration of adaptive compression frameworks together with heterogeneous datasets characterized by multi-modal inputs, i.e., images, time series, or text data, to fully characterize the energy demands of FL processes and identify the optimal operating conditions.